# Evaluation at the Remote Site for Ultra-stable Radio Frequency Dissemination via Fiber Links

Shanglin Li, Haoyuan Lu, Shuangyou Zhang, Dawei Li, Jianxiao Leng and Jianye Zhao

*Abstract*—We demonstrate a method which can directly evaluate the radio frequency transfer quality via fiber links at the remote site. Coherent signals are first transferred to the same remote site via two stabilized fiber links. The two signals at the remote site are compared with each other. The relative phase difference can represent transfer stability loss. This evaluation method at the remote site has been compared with the traditional one with which the signal is evaluated at the local site. The two results match perfectly. It indicates that the method is available to evaluate the transfer performance of radio frequency (RF) dissemination in such applications as antenna array systems.

*Index Terms*—Optical fiber link, radio frequency dissemination, antenna array systems.

## I. INTRODUCTION

FREQUENCY signals can be disseminated over the optical fiber, which benefits modern science and practical applications, such as the comparison of optical clocks, fundamental physics measurement and antenna arrays [1-4]. Fiber links always suffer from phase noise caused from temperature fluctuation and mechanical vibration [5]. Therefore, it is necessary to evaluate the frequency transfer quality via fiber links. In the previous experiments [5-9], the signal at the remote site has to be compared with the reference signal at the local site to evaluate the performance of the transfer system.

More and more multi-antenna systems are constructed for space navigation and tracking, such as NASA Deep Space Communications Complex (DSCC) [10] and Square Kilometer Array (SKA) project [11]. Precise and highly coherent frequency signals need to be delivered stably from the signal processing center to all the remote antennas through fiber links for multiple astronomy observation missions [12]. In Square Kilometer Array (SKA) project, complex fiber networks which link the signal processing center and antenna arrays sites are being built [13]. High-precision and coherent frequency signal will be transferred to the antenna arrays site via networks. With some real-time tasks undertaken such as real-time imaging, transfer quality from the central station to each antenna also needs to be evaluated in real time [11]. Multi-antenna systems such as SKA need to transmit and process vast amounts of data, occupying vast transportation network and consuming immense computing power and large storage space. When the frequency signal at the antenna array side is not stable enough, data transmission is unnecessary. The traditional method, with which the frequency signal is evaluated at the local site, means that the signal has to be transferred back to the local site for evaluation. And the round-trip time delay will cause the large waste of the relevant resource. If we evaluate the transfer performance directly at the antenna array side, backward transfer delay is removed and we will control the data transmission in real time. Therefore, it is necessary to find an effective way to evaluate the transfer performance directly at the antenna array side.

Compared with the experiment in this paper, the previous experiment demonstrated an antiparallel configuration to evaluate the performance of optical carrier frequency dissemination system [14]. But it is not fit for RF transfer between the signal processing center and the antenna array side for following reasons. When we use the similar method, an ultrastable radio frequency standard is needed as the reference of the counter. Besides, if the timing jitter reaches the order of magnitude of $10^{-15}$[15], it is difficult for the commercial frequency counters to satisfy our demands. Moreover, the antiparallel configuration requires the existence of the frequency standard at the both sites, however the standard only exists at the signal processing center in SKA. In this case, the antiparallel configuration is not suitable.

In this paper, we demonstrate a method to solve the problem which is proposed above.

## II. THEORY AND PRINCIPLES

In Fig.1 (a), it shows a traditional way to evaluate the performance of radio frequency transfer. At the local site, the frequency signal is injected into the fiber. When the signal reaches the remote site, one part is retro-reflected to the local site for compensating the phase fluctuation induced by the environment in the fiber link, while the rest is detected at the remote site. In order to evaluate the transfer quality, the signal transferred is compared with the reference source at the local site. In other words, the remote site and the local site have to be located at the same place.

In Fig.1 (b), a method that can evaluate the transfer performance directly at the remote site is illustrated. Coherent frequency signals at the local site are disseminated to the remote site via two different fiber links, and the two fiber links are stabilized respectively according to the compensation module. At the remote site, two signals are detected and are compared to get the beat note. With this link configuration, the

This work is supported by National Natural Science Foundations of China (No. 61371074). (Corresponding author: Jianye Zhao.)
The authors are with Department of Electronics, School of Electronics Engineering and Computer Science, Peking University, Beijing, 100871, China. (e-mail: zhaojianye@pku.edu.cn).

uncertainty of the frequency source is cancelled, and the phase noise of both fibers is mixed. Hence, an upper limit for the residual instability of the disseminated RF signal is acquired. The upper limit can be used to evaluate the relative transfer performance of fiber links.

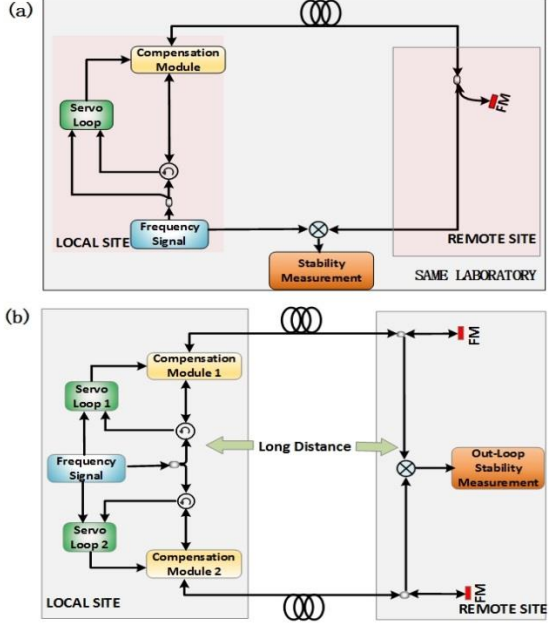

Fig. 1. The comparison of two evaluation methods for RF transfer. Fig.1(a) represents the traditional evaluation method in which the local site and the remote site must be in the same place; Fig.1(b) represents the evaluation method that can evaluate the transfer performance directly at the remote site. PD: photodiode, FM: Faraday Mirror.

As the Fig.2 shows, two RF signals mix with each other to derive the stability loss at the remote site. In addition, we also use the traditional evaluation method to obtain the stability loss of fiber link 1 and 2 respectively. When the stability loss of two fiber links is different, the better one will act as the reference. And the stability loss measured at the remote site should match the worse stability loss of fiber link 1 and 2.

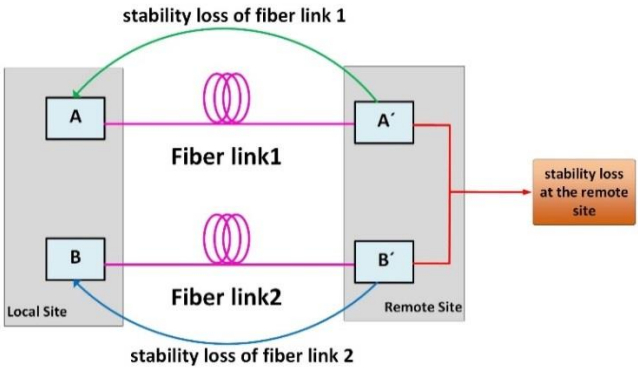

Fig. 2. The red line shows that the signals are mixed at the remote site and utilized to obtain stability loss. The green line shows that the stability loss of fiber link 1 is derived with the traditional evaluation method. The blue line shows that the stability loss of fiber link 2 is derived with the traditional evaluation method.

### III. EXPERIMENTAL SETUP AND THE RESULTS

Experiments are carried out over single-mode fiber (SMF) links. As a test of phase-difference measurements, we link the local site and the remote one via fiber spools which are located at different positions of the same place. The optical frequency comb is capable of providing microwave signals at a noise below the shot noise of light [16], so in our experiment, we use optical frequency comb which is generated from a mode-locked laser as RF source. The comb's repetition frequency is 100 MHz. The frequency of the first, second, and third harmonic is 100MHz, 200MHz, 300MHz respectively. In Fig.3, signals are transferred over 20-km fiber spools. The coherent frequency signals are obtained by extracting the harmonics of the repetition frequency of the frequency comb, then utilized as the references of two independent frequency transfer system. For the first fiber link, phase compensation scheme consists of an electrical mixer which mixes the retro-reflected signal with the third harmonic, and the output signal is transferred to the remote site. In the second fiber link, a phase shifter and a servo loop are involved to actively compensate the phase fluctuation.

The output of an Er-doped mode-locked laser is detected by a photodiode (PD) to acquire RF spectrum which consists of a group of coherent frequencies. Then band pass filters (BPFs) are applied to filter out the harmonics.

For fiber link 1, the first harmonic signal, $V_{1,1}(t) = A_{1,1}\cos(\omega t + \varphi_1)$ is transferred to the remote site and retro-reflected back to the local site [9]. Here $\omega$ equals the baseband frequency and $\varphi_1$ is the initial phase. Resulting from vibrations and temperature fluctuation in the fiber link, phase noise $\varphi_{phase1}$ is induced. The round-trip signal $V_{1,retro}(t) = A_{1,retro}\cos[\omega(t - 2\tau_1) + \varphi_1 - 2\varphi_{phase1}]$ is phase-conjugated with the third harmonic $V_{1,3}(t) = A_{1,3}\cos(3\omega t + \varphi_1 + \varphi_d)$. $\tau_1$ represents the one-way time delay and $\varphi_d$ represents the fixed initial phase difference between the first harmonic and the third one. The signal $V_{1,2}(t)$ after phase conjugation can be expressed as $A_{1,2}\cos[2\omega(t + \tau_1) + \varphi_d + 2\varphi_{phase1}]$, which is transferred to the remote site. As a result, at the remote site, the signal $V_{1,remote}(t) = A_{1,remote}\cos(2\omega t + \varphi_d)$ is theoretically immune to phase noise induced from vibrations and temperature fluctuation.

For fiber link 2, the transferred signal is the first harmonic signal $V_{2,1}(t) = A_{2,1}\cos(\omega t + \varphi_1)$. After a phase shifter, the signal is changed into $V_{2,c}(t) = A_{2,c}\cos(\omega t + \varphi_1 + \varphi_c)$, where an extra phase $\varphi_c$ is used to actively compensate the phase fluctuation [17]. The signal at the remote site $V_{2,4}(t) = A_{2,4}\cos[\omega(t - \tau_2) + \varphi_1 + \varphi_c - \varphi_{phase2}]$ is divided into two parts. $\varphi_{phase2}$ represents the phase fluctuation and $\tau_2$ represents one-way time delay over fiber link 2. The major part is sent back for actively compensating the phase fluctuation. The round-trip signal can be expressed as $V_{2,retro}(t) = A_{2,retro}\cos[\omega(t - 2\tau_2) + \varphi_1 + \varphi_c - 2\varphi_{phase2}]$. It mixes with the phase-shifted signal $V_{2,c}(t)$, and the output signal, processed by a band pass filter, is $V_{2,2}(t) = A_{2,2}\cos[2\omega(t - \tau_2) + 2\varphi_1 + 2\varphi_c - 2\varphi_{phase2}]$. The error signal $V_{2,err}(t) = A_{2,err}\cos[2\varphi_c - 2\varphi_{phase2} - 2\omega\tau_2]$ is acquired by mixing $V_{2,2}(t)$ with $V_{2,3}(t)$. It is used to control the phase shifter to suppress the phase fluctuation by keeping $\varphi_c - \varphi_{phase2}$ constant. As a result, the signal $V_{2,4}(t) = A_{2,4}\cos[\omega(t - \tau_2) +$

$\varphi_1$] is also immune to phase noise. In order to evaluate the relative stability loss, the frequency of $V_{2,4}(t)$ is doubled to $V_{2,remote}(t)$.

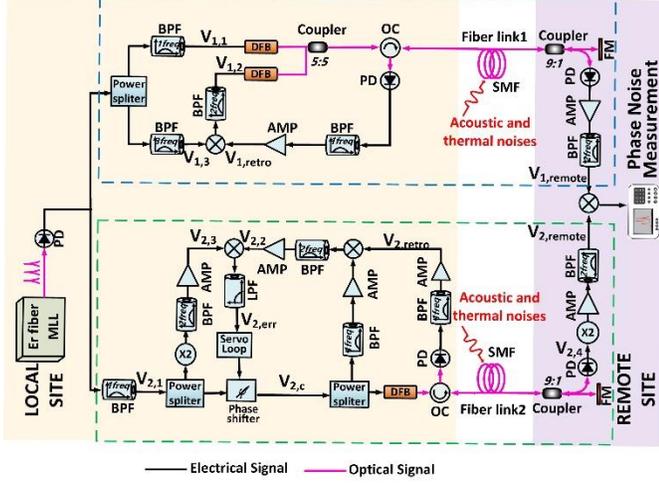

Fig. 3. The schematic of the experiment over 20 km fiber links. BPF: band pass filter, LPF: low pass filter, DBF: distribute feedback laser, AMP: amplifier, OC: optical circulator, PD: photodiode, FM: Faraday mirror, SMF: single mode fiber

Adopting the method illustrated in Fig.1 (b), we mix $V_{1,remote}(t)$ with $V_{2,remote}(t)$ at the remote site to evaluate the relative transfer performance of fiber links. A mixer multiplies the input signals and the component of high frequency can be easily removed with a low pass filter. The phase difference of the two signals is detected in term of voltage fluctuation and recorded by a high resolution voltage acquisition instrument (Agilent 34401A). Meanwhile, we derive the stability loss of two fiber links respectively using the traditional evaluation method and the noise floor. We use different compensation modules over two fiber links in our experiment to completely ensure the system noises are different, because the same system noises might be inter-neutralized during the phase comparison progress to influence the accuracy of the evaluation.

The transfer performances are expressed by Overlapping Allan Deviation illustrated in Fig.4. The relative stability loss we measure directly at the remote site is $8 \times 10^{-14}$ at 1 s, $2.5 \times 10^{-15}$ at 100 s and $6 \times 10^{-16}$ at 1000 s. In order to assess the rationality of our method, stability loss of both fiber links is measured with the traditional evaluation method respectively. The stability loss of fiber link 1 is $8 \times 10^{-14}$ at 1s, $2.2 \times 10^{-15}$ at 100 s and $5 \times 10^{-16}$ at 1000 s, while the stability loss of fiber link 2 is $5 \times 10^{-14}$ at 1s, $1.5 \times 10^{-15}$ at 100 s and $2 \times 10^{-16}$ at 1000 s. It shows the compensation schemes is effective compared with free running performance. The residual frequency instability is $2 \times 10^{-13}$ at 1 s, $3 \times 10^{-15}$ at 1000 s and $4 \times 10^{-15}$ at 4000s for the free running system. In addition, noise floor of the system shown as the black curve is $4.8 \times 10^{-14}$ at 1s, $9 \times 10^{-17}$ at 1000s and $3 \times 10^{-17}$ at 4000s.

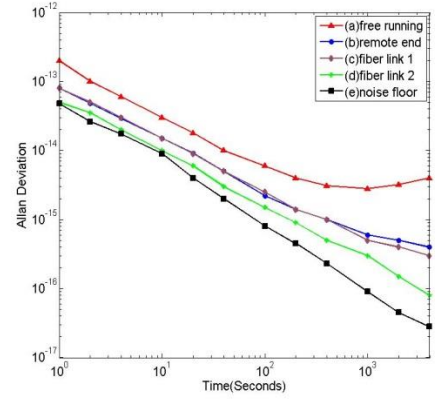

Fig. 4. Frequency stability loss measurements over 20 km fiber links. The curve (a) shows the stability loss for the free running system; The curve (b) shows relative stability loss which is measured directly at the remote site; The curve (c) and curve (d) show the stability loss of fiber link 1 and 2 which are measured respectively with the traditional method; The curve (e) shows the phase noise floor of the system.

Besides, we do the experiment where the round-trip distance of fiber is 100 km. We use EDFA to amplify the optical power, because the round-trip distance is about 100km and the power is very weak. The schematic of the experiment is illustrated in Fig.5.

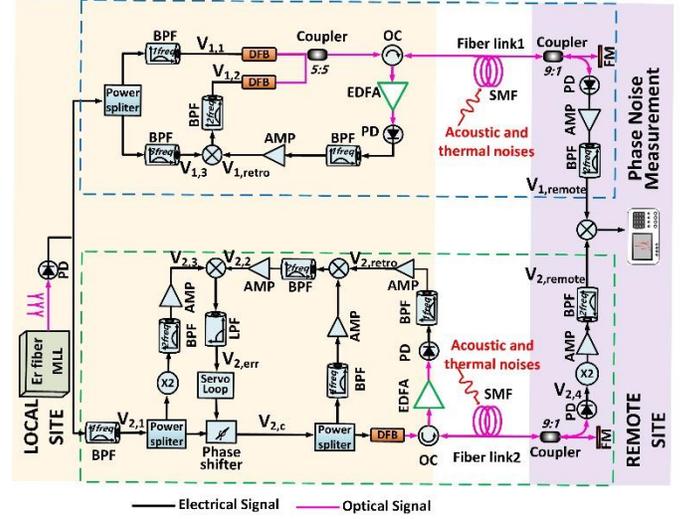

Fig. 5. The schematic of the experiment where round-trip distance is 100km. BPF: band pass filter, LPF: low pass filter, DBF: distribute feedback laser, AMP: amplifier, OC: optical circulator, PD: photodiode, FM: Faraday mirror, SMF: single mode fiber, EDFA: Erbium Doped Fiber Amplifier.

The transfer performances are illustrated in Fig.6. The relative stability loss we measure directly at the remote site is $3 \times 10^{-13}$ at 1 s, $2.5 \times 10^{-14}$ at 100 s and $6.7 \times 10^{-15}$ at 1000 s. The stability loss of fiber link 1 is $2.9 \times 10^{-13}$ at 1s, $2.5 \times 10^{-14}$ at 100 s and $6.2 \times 10^{-15}$ at 1000 s, while the stability loss of fiber link 2 is $9.5 \times 10^{-14}$ at 1s, $2.5 \times 10^{-15}$ at 100 s and $4.8 \times 10^{-16}$ at 1000 s. It shows the compensation scheme is effective compared with free running performance. Besides, the residual frequency instability is $7 \times 10^{-13}$ at 1 s, $1.3 \times 10^{-14}$ at 1000 s and $1.7 \times 10^{-14}$ at 4000s for the free running system.

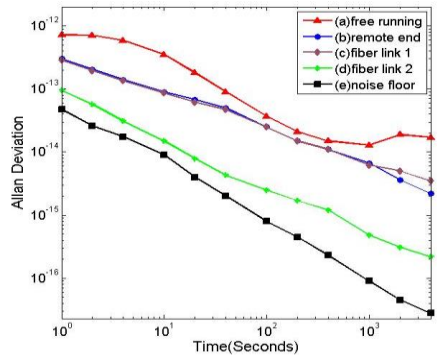

Fig. 6 Frequency stability loss measurements for the round-trip link, 100km. The curve (a) shows the stability loss for the free running system; The curve (b) shows relative stability loss which is measured directly at the remote site; The curve (c) and curve (d) show the stability loss of fiber link 1 and 2 which are measured respectively with the traditional method; The curve (e) shows the phase noise floor of the system.

## IV. CONCLUSION

According to the discussion above, if two signals mix with each other at the remote site, the Allan deviation we derive indicates an upper limit for the residual instability of the disseminated frequency signal, which can be used to evaluate the relative transfer performance of fiber links. When the stability loss of two fiber links is different, the better one will act as the reference in fact. In the experiment above, the signal transferred by fiber link 2 serves as the reference, and the relative stability loss derived directly at the remote site will be more close to that of fiber link 1. In fact, the stability loss measured at the remote site shares almost the same curve with the stability loss of fiber link 1. The result confirms the previous analysis. Hence, we can draw a conclusion: it is reasonable to use the relative stability loss of two independent fiber links to evaluate the RF transfer performance directly at the remote site. Our method removes the backward transfer time delay and can be used to control the data transmission in real time. It will reduce the needless consumption of transportation network resource, computing power and storage space in antenna array systems such as SKA. With our method, we can measure the relative stability for the RF signal transfer between the processing center and the remote antennas and evaluate the transfer performance directly at the antenna array side.

## REFERENCES

[1] Rosenband, T., Hume, D. B., Schmidt, P. O., Chou, C. W., Brusch, A., Lorini, L., ... , Diddams. "Frequency ratio of Al+ and Hg+ single-ion optical clocks; metrology at the 17th decimal place," *Science*, vol. 319, no. 5871, pp. 1808-1812, Mar, 2008.
[2] Arthur Matveev, Christian G. Parthey, Katharina Predehl, Janis Alnis, Axel Beyer, Ronald Holzwarth, Thomas Udem, Tobias Wilken, Nikolai Kolachevsky, Michel Abgrall, Daniele Rovera, Christophe Salomon, Philippe Laurent, Gesine Grosche, Osama Terra, Thomas Legero, Harald Schnatz, Stefan Weyers, Brett Altschul, and Theodor W. Hansch, "Precision Measurement of the Hydrogen 1S-2S Frequency via a 920-km Fiber Link," *Phys. Rev. Lett*, vol. 110, 230801, Jun, 2013.
[3] F.-L. Hong, M. Musha, M. Takamoto, H. Inaba, S. Yanagimachi, A. Takamizawa, K. Watabe, T. Ikegami, M. Imae, Y. Fujii, M. Amemiya, K. Nakagawa, K. Ueda, and H. Katori, "Measuring the frequency of a Sr optical lattice clock using a 120 km coherent optical transfer," *Opt. Lett.*, vol. 34, no. 5, pp. 692–694, Feb, 2009.
[4] S. M. Foreman, K. W. Holman, D. D. Hudson, D. J. Jones, and J. Ye, "Remote transfer of ultrastable frequency references via fiber networks," *Rev. Sci. Instrum*, vol. 78, no. 2, 021101, Feb, 2007.
[5] S. Droste, F. Ozimek, Th. Udem, K. Predehl, T. W. Hansch, H. Schnatz, G. Grosche, and R. Holzwarth, "Optical-Frequency Transfer over a Single Span 1840 km Fiber Link," *Phys. Rev. Lett,* vol. 111, 110801, May,2013.
[6] Miho Fujieda, Motohiro Kumagai, Shigeo Nagano, Atsushi Yamaguchi, Hidekazu Hachisu, and Tetsuya Ido, "All-optical link for direct comparison of distant optical clocks," *Opt. Exp.*, vol. 19, no. 17, pp. 16498–16507, Aug, 2011.
[7] Olivier Lopez, Adil Haboucha, Bruno Chanteau, Christian Chardonnet, Anne Amy-Klein, and Giorgio Santarelli, "Ultra-stable long distance optical frequency distribution using the Internet fiber network," *Opt. Exp.*, vol. 20, no. 21, pp. 23518–23526, Oct, 2012.
[8] D. Calonico, E. K. Bertacco, C. E. Calosso, C. Clivati, G. A. Costanzo, M. Frittelli, A. Godone, A. Mura, N. Poli, D. V. Sutyrin, G. Tino, M. E. Zucco, F. Levi, "High-accuracy coherent optical frequency transfer over a doubled 642-km fiber link," *Appl. Phys. B*, vol. 117, no. 3, 979, Sep, 2014.
[9] Dawei Li, Dong Hou, Ermeng Hu, and Jianye Zhao, "Phase conjugation frequency dissemination based on harmonics of optical comb at $10^{-17}$ instability level," *Opt. Lett.*, vol. 39, no. 17, pp. 5058–5061, Aug, 2014.
[10] Calhoun M, Huang S, Tjoelker R L, "Stable Photonic Links for Frequency and Time Transfer in the Deep-Space Network and Antenna Arrays," *Proceedings of the IEEE*, vol. 95, pp. 1931-1946, Oct. 2007
[11] Wang B, Zhu X, Gao C, Bai Y, Dong JW, Wang LJ, "Square Kilometre Array Telescope—Precision Reference Frequency Synchronisation via 1f-2f Dissemination," *Sci. Rep.*, Sept. 2015.
[12] S. Huang and R. Tjoelker, "Stabilized photonic link for frequency andtime transfer in antenna arrays," presented at Proc. 38th PTTI, Dec. 7-9, 2006.
[13] Mccool R, Garrington S, Spencer R., 'Signal transport and networks for the SKA", General Assembly and Scientific Symposium, Istanbul, Turkey , 2011, pp. 1 – 4.
[14] K. Predeh, G. Grosche, S. M. F. Raupach, S. Droste, O. Terra, J. Alnis, Th. Legero, T. W. Hänsch, Th. Udem, R. Holzwarth, and H. Schnatz, "A 920-Kilometer Optical Fiber Link for Frequency Metrology at the 19th Decimal Place," *Science,* vol. 336, no. 6080, pp. 441-444, Apr, 2012.
[15] Kwangyun Jung, Junho Shin, Jinho Kang, Stephan Hunziker, Chang-Ki Min, and Jungwon Kim, "Frequency comb-based microwave transfer over fiber with 7×10$^{-19}$ instability using fiber-loop optical-microwave phase detectors," *Opt Lett,* vol. 39, no. 6, pp. 1577-1580, Mar, 2014.
[16] T. M. Fortier, M. S. Kirchner, F. Quinlan, J. Taylor, J. C. Bergquist, T. Rosenband, N. Lemke,A. Ludlow, Y. Jiang, C. W. Oates , S. A. Diddams, "Generation of ultrastable microwaves via optical frequency division," Nat. Photonics, vol. 5, no. 7, pp. 425-429, Jun, 2011.
[17] Jianye Zhao, Dawei Li, Bo Ning, Shuangyou Zhang, and Wei Duan," Highly-stable frequency transfer via fiber link with improved electrical error signal extraction and compensation scheme," *Opt. Exp.*, vol. 23, no. 7, pp. 8829-8836, Apr, 2015.